\documentclass[aps,pra,preprint,superscriptaddress,showpacs]{revtex4-1}
\usepackage{graphicx}
\usepackage[utf8]{inputenc}

\begin{document}


\title{Continuous spectra in high-harmonic generation driven by multicycle laser pulses}

\author{Warein Holgado}
\email[]{warein@usal.es}
\affiliation{Grupo de Investigaci\'on en \'Optica Extrema (GIOE), Universidad de Salamanca, Pl. de la Merced s/n, E-37008 Salamanca, Spain}

\author{C. Hern\'andez-Garc\'ia}
\affiliation{Grupo de Investigaci\'on en \'Optica Extrema (GIOE), Universidad de Salamanca, Pl. de la Merced s/n, E-37008 Salamanca, Spain}
\affiliation{JILA, University of Colorado at Boulder, 440 UCB Boulder, CO 80309-0440, USA}

\author{B. Alonso}
\affiliation{Grupo de Investigaci\'on en \'Optica Extrema (GIOE), Universidad de Salamanca, Pl. de la Merced s/n, E-37008 Salamanca, Spain}
\affiliation{IFIMUP-IN and Departamento de F\'isica e Astronomia, Faculdade de Ciências, Universidade do Porto, Rua do Campo Alegre 687, 4169-007 Porto, Portugal}

\author{M. Miranda}
\affiliation{IFIMUP-IN and Departamento de F\'isica e Astronomia, Faculdade de Ciências, Universidade do Porto, Rua do Campo Alegre 687, 4169-007 Porto, Portugal}
\affiliation{Department of Physics, Lund University, P.O. Box 118, SE-221 00 Lund, Sweden}

\author{F. Silva}
\affiliation{IFIMUP-IN and Departamento de F\'isica e Astronomia, Faculdade de Ciências, Universidade do Porto, Rua do Campo Alegre 687, 4169-007 Porto, Portugal}

\author{L. Plaja}
\affiliation{Grupo de Investigaci\'on en \'Optica Extrema (GIOE), Universidad de Salamanca, Pl. de la Merced s/n, E-37008 Salamanca, Spain}

\author{H. Crespo}
\affiliation{IFIMUP-IN and Departamento de F\'isica e Astronomia, Faculdade de Ciências, Universidade do Porto, Rua do Campo Alegre 687, 4169-007 Porto, Portugal}

\author{I. J. Sola}
\affiliation{Grupo de Investigaci\'on en \'Optica Extrema (GIOE), Universidad de Salamanca, Pl. de la Merced s/n, E-37008 Salamanca, Spain}


\begin{abstract}
	We present observations of the emission of XUV continua in the 20-37 eV region by high harmonic generation (HHG) with $4$-$7\ \mathrm{fs}$ pulses focused onto a Kr gas jet. The underlying mechanism relies on coherent control of the relative delays and phases between individually generated attosecond pulse, achievable by adjusting the chirp of the driving pulses and the interaction geometry. Under adequate negative chirp and phase matching conditions, the resulting interpulse interference yields a continuum XUV spectrum, which is due to both microscopic and macroscopic (propagation) contributions. This technique opens the route for modifying the phase of individual attosecond pulses and for the coherent synthesis of XUV continua from multicycle driving laser pulses without the need of an isolated attosecond burst.
\end{abstract}

\pacs{42.65.Ky, 42.65.Re}

\maketitle

\section{Introduction}
High-order harmonic generation (HHG) is a process where, from the nonlinear interaction of an intense laser pulse with an atom or a molecule, new frequencies in the extreme ultraviolet (XUV) and soft-x-ray range are generated. This can be described by a semi-classical three-step model \cite{corkum1993,schafer1993}, and the radiation is emitted as a train of bursts with durations down to the attosecond scale, commonly presenting an odd harmonic comb-like spectrum.

Since the advent of HHG and subsequently of attosecond science, the generation of continua in the XUV range has been of high interest. This has firstly been associated to the process of emission of a single attosecond light pulse \cite{hentschel2001,sansone2006}, but it also presents interest as a broadband source for spectroscopic applications. For instance, characterization of the optical properties of materials and components in the XUV range is challenging, usually requiring synchrotron facilities for broadband spectral measurements. Recently, typical comb-like HHG spectra have also proved to be a tool for XUV optical characterization. For instance, Goh and coworkers have applied a HHG source for diffraction grating characterization \cite{goh2015}. An HHG source has also been applied to mirror reflectance analysis, complementing synchrotron source assisted measurements of reflectivity with spectral phase characterization by employing the odd-harmonic comb-like spectrum from HHG and an attosecond pulse reconstruction technique \cite{anne2006}. In this case, the comb-like spectrum provides access to the phase information at some wavelengths (the odd harmonics) but, in this context, a continuous coherent XUV spectrum may represent an interesting source capable of accessing a broader and more complete spectral region. Such spectra are foreseen also as a key for extending spectral interferometric techniques, such as optical coherence tomography, to the XUV domain \cite{fuchs2012}. Unlike comb-like spectra, continuous coherent XUV spectra enable easier observation of spectral interference over a continuous range and the coherence of HHG radiation allows for the measurement of not only amplitude changes but also of phase variations during a certain experiment or material interaction. Therefore, a coherent continuum XUV source enables a more complete characterization (amplitude and phase) in this spectral range. As a last example, the powerful techniques of time-resolved and transient absorption XUV spectroscopy are behind very active research aimed at understanding the electron dynamics in atoms and molecules (see, e.g., \cite{holler2011,chini2012,ott2014}) although the use of a probe beam with a comb-like spectra from HHG limits the analyzed region to that covered by the harmonics. Goulielmakis \emph{et al} have shown that by using an isolated attosecond pulse the technique can give more information, mainly due to the fact that it corresponds to a continuum in the spectral domain \cite{gouliel2010}.


To date, most of the strategies to obtain XUV continua have relied on isolating an attosecond pulse at the microscopic level (i.e. single atom). This implies the occurrence of a single ionization-flight-recombination event, which is only possible if the driving field pulse is as short as the period of the central frequency or, in the case of few-cycle pulses, if HHG is confined by gating techniques. Different strategies have been proposed, based on carrier-envelope phase stabilized few-cycle driving fields, namely spectral selection of the cutoff \cite{hentschel2001}, polarization gating \cite{sansone2006,sola2006} and double optical gating \cite{chang2007}, among others. The drawback of such techniques is the decrease of the generation efficiency, making it difficult to apply them to low-energy driving fields, such as the new high repetition rate $\mu$J-level OPCPA sources.\cite{krebs2013,rudawski2015}. Macroscopic phenomena can also be relevant to isolate attosecond pulses \cite{mairesse2004}. For instance, different techniques that make use of phase-matching conditions have been proposed, such as the use of flat-top beams at the focus \cite{strelkov2008}, ionization gating \cite{thomann2009,ferrari2010}, or tilted pulse beams \cite{wheeler2012}. Recently, phase matching on high pressure gases has been proposed to obtain a single XUV burst \cite{chen2014}. However, all the mentioned techniques generate XUV continua through the emission of an isolated attosecond pulse.

In previous works, the effect of the chirp of the driving pulse over HHG generated on gas targets, which is directly related to the delay between each attosecond burst, has been studied  \cite{chang1998,zhou1996}. The continuous spectrum emerged in the cutoff region for IR pulses that were negatively or highly positively chirped, while the plateau remained insensible to changes in the IR chirp. On the other hand, Lee \emph{et al.} measured continuous spectra in HHG performed in Neon \cite{lee2001} for harmonic orders between 60 and 90 (61-136 eV) and driving pulses that were negatively and positively chirped.

In this work we explore the generation of XUV continua driven by multi-cycle laser pulses, by altering the relative delay and phase of a train of attosecond bursts, without going through the complexities of generating an isolated pulse. We present experimental results, corroborated by theoretical simulations, that demonstrate the emission of XUV continua ranging from 20 to $37\ \mathrm{eV}$ as a result of the combination of microscopic and macroscopic physics. In contrast to previous works, the generated XUV continua lie in the low HHG \emph{plateau} (wavelengths between $30$ and $60\ \mathrm{nm}$). Furthermore, we present theoretical results that prove the influence of propagation in the resulting spectra.

In the first part of this paper we present the experimental setup and results. In the second part, theoretical simulations are presented, discussed and compared to experiments, whereby the nature of the obtained XUV continua is inferred.

\section{Experimental Results}
\label{sec:examples}

In the experiments we use a $1$-kHz Ti:Sapphire CPA amplifier (FemtoPower CompactPRO CE-Phase, Femtolasers GmbH) delivering pulses with a Fourier-transform limited duration of $25\ \mathrm{fs}$ (FWHM) and energy up to $0.9\ \mathrm{mJ}$. The output pulses are spectrally broadened in a hollow-core fiber (HCF) with an inner diameter of $250\ \mu\mathrm{m}$ and $1\ \mathrm{m}$  length. The HCF is filled with argon at $1\ \mathrm{bar}$ pressure. The carrier-envelope phase (CEP) is stabilized in the oscillator, with a CEP RMS of 100 mrad in the fast loop, and well-preserved during the amplification and post-compression processes.

The resulting broadband pulses are then sent through a chirped mirror and glass wedge compressor that forms part of a dispersion-scan (d-scan) \cite{miranda2012} pulse measurement and compression system. By compensating the spectral phase with 10 bounces off double-angle chirped mirrors (Ultrafast Innovations GmbH; nominal GDD: $-40\ \mathrm{fs}^2$ per pair at $800\ \mathrm{nm}$, minimum reflectance: $99$\%) and addition of normal dispersive material (BK7 wedge pair), few-cycle pulses are obtained. Using this setup we can routinely achieve pulse durations down to sub-$4\ \mathrm{fs}$ \cite{alonso2013,silva2014}. Depending on the gas pressure, pulse duration can be tuned in a certain range. In our experiments we work with pulse durations from $4$ to $7\ \mathrm{fs}$ (FWHM), as determined with d-scan measurements.

The laser pulses are then focused by a spherical silver mirror ($f=500\  \mathrm{mm}$) into a pulsed Kr gas jet, with an estimated intensity on target of $2\times10^{14}\ \mathrm{W/cm}^2$. Since krypton has lower $I_p$ than argon, we use Kr to obtain a larger HHG signal \cite{liang1997}. The spherical mirror is placed on a translation stage, so the position of the focus can be controlled. The pulse enters the vacuum chamber through a $0.5\ \mathrm{mm}$ thick fused-silica window, situated close to the focusing mirror to avoid possible nonlinear effects. HHG is driven in a Kr gas jet ($5\ \mathrm{bar}$ of backing pressure) using a nozzle of $500\ \mu\mathrm{m}$ diameter. The pressure reached inside the vacuum chamber where the high-order harmonics are generated is around $5\times 10^{-3}\ \mathrm{mbar}$. A $150$-$\mathrm{nm}$ thick aluminum foil is used to filter the IR radiation and the lower harmonics. The HHG spectra are characterized with a grazing-incidence Rowland circle XUV spectrometer (Model 248/310G, McPherson Inc.), of $1$-$\mathrm{m}$ radius and $133$-grooves/mm spherical diffraction grating.

\begin{figure}[h]
	\centering
	\includegraphics[scale=0.5]{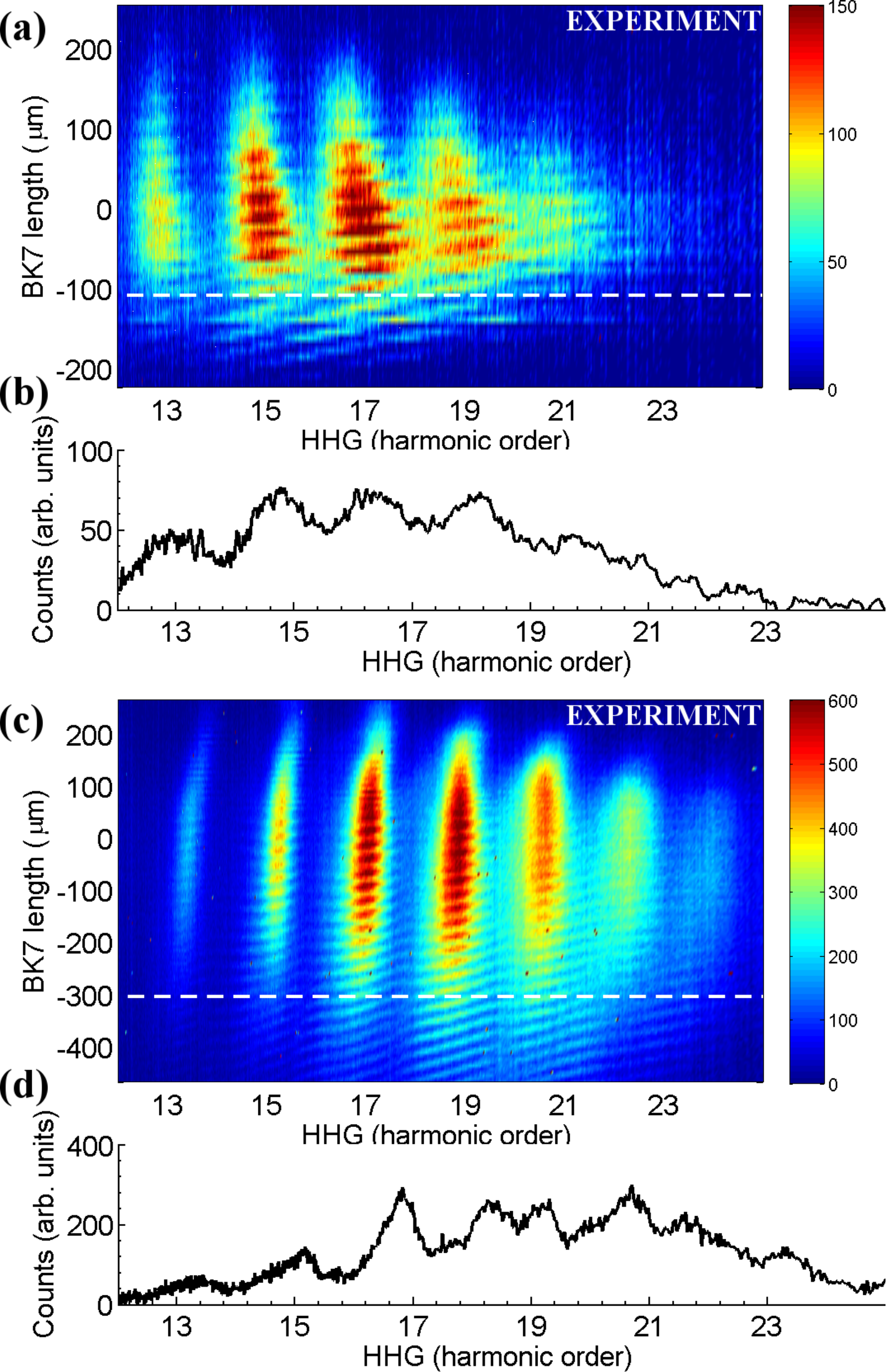}
	\caption{Measured HHG dependence on BK7 insertion for $4\ \mathrm{fs}$ pulses (a) and $7\ \mathrm{fs}$ pulses (c). When the insertion is finely varied, CEP dependence can be seen in the HHG spectra in the form of fringes. Below each scan, in (b) and (d), XUV spectra are shown for the negative chirp BK7 insertion shown as a white dashed line in the corresponding chirp-scan.}
	\label{fig1}
\end{figure}

The gas jet is placed $4\ \mathrm{mm}$ and $2\ \mathrm{mm}$ before the focus position for $4\ \mathrm{fs}$ and $7\ \mathrm{fs}$ pulses respectively. Gas-jet is placed before the focus to select favorable transversal phase-matching conditions for the observation of CEP-dependent HHG \cite{carlos2015}. A dispersion scan is performed on the driving pulses by finely translating one of the BK7 wedges. Therefore, HHG is studied as a function of the spectral phase of the incident beam. In Fig.$\ $\ref{fig1} we present the variation of the HHG spectra as a function  of the BK7 insertion. The map reveals how crucial the effect of the IR spectral phase on the HHG is. As depicted, the narrower structures appear when the generating beam is positively chirped, while at negative chirping the spectra broaden, until eventually becoming continua covering over 12 harmonic orders (around $17\ \mathrm{eV}$). The observed fringes denote the CEP dependence of the HHG (i.e. for a short range of BK7 insertion, the dispersion-scan becomes a CEP scan).

In Fig.$\ $\ref{fig1}(c) we show the CEP-scan for a 7 fs driving pulse. The range of dispersion values generating an efficient HHG signal is narrower for $4\ \mathrm{fs}$ pulses [Fig.$\ $\ref{fig1}(a)] than for $7\ \mathrm{fs}$ pulses, since the spectral content of the former is broader than that of the latter, but in both cases the continuum appears for negative chirp [Figs.$\ $\ref{fig1}(b) and \ref{fig1}(d)]. 


In Fig.$\ $\ref{fig1}(b) we show the continuum achieved with $4\ \mathrm{fs}$ laser pulses for a BK7 insertion of $-108\ \mu\mathrm{m}$, i.e., when $108\ \mu\mathrm{m}$ are removed from the Fouier limit condition. In order to quantify the quality/smoothness of the achieved continuum, we calculate the corresponding contrast for each harmonic order. Contrast is defined as
\begin{equation}
C=\frac{I_{max}-I_{min}}{I_{max}+I_{min}}.
\end{equation}

We compute this value for regions of two orders of harmonic. The mean contrast in the case of $4\ \mathrm{fs}$, in a range from the 13th to the 21st harmonics is $0.43$. In the case of $7\ \mathrm{fs}$ laser pulses, for a BK7 insertion of $-303\ \mu\mathrm{m}$ [shown in Fig.$\ $\ref{fig1}(d)] and a spectrum between harmonic orders 15th and 23th, the mean contrast is $0.40$. Even though these are not totally smooth (unmodulated) continua, there are no wavelengths with zero signal in this range.

Contrarly, in the case of positive chirp, these values are $0.62$ for $4\ \mathrm{fs}$ (insertion of $100\ \mu\mathrm{m}$) and $0.76$ for $7\ \mathrm{fs}$ (insertion of $152\ \mu\mathrm{m}$).


It is already known that HHG shows a spectral broadening/narrowing when negative/positive IR pulses are employed \cite{chang1998,zhou1996,p1997,lee2001}, contrarily to what occurs in the self-phase modulation process \cite{p1997}. However, the continua described in those works were only achieved for very high harmonic orders (in Ne, $>70\ \mathrm{eV}$).  Calegari et al.$\ $\cite{calegari2011} reported continua in Xe using $5\ \mathrm{fs}$ pulses focused into a gas cell in a high intensity regime ($2.5\times10^{15}\ \mathrm{W/cm}^2$) when selecting the so-called short quantum paths. According to the authors, this effect is introduced by the plasma-induced chirp of the driving field as one main factor, yielding a single attosecond pulse in the time domain. In our case, we perform a scan over chirp to fully understand its effect.

\section{Theoretical simulations}

In the observed case, continua generated for negatively chirped driving pulse conditions appear dramatically in a relatively low energy region ($20$-$37\ \mathrm{eV}$), covering both the HHG plateau and the cutoff. To our knowledge, to date only one other similar behavior has been observed \cite{rudawski2015}. 

The appearance of continua when using negatively chirped driving pulses is reproduced by our calculations. We compute harmonic propagation using a method based on the electromagnetic field propagator \cite{carlos2010}. The dipole acceleration of each elementary source is computed using the SFA+ method \cite{jose2009}, an extension of the strong field approximation. Due to the lack of contrasted Coulomb potentials for krypton, we perform our HHG calculations in argon (which has similar ionization potential), using the Coulomb potential given in \cite{tong2005}. Propagation effects of the fundamental field, including plasma and neutral dispersion as well as time-dependent group velocity walk-off, are all taken into account. The absorption of the harmonics in the gas is modeled using Beer's law. Agreement between the model and theory when phase matching is a relevant factor was already tested in Refs. \cite{chen2014,carlos2015,carlos2013,kretschmar2013}. 

\begin{figure}[b]
	\centering
	\includegraphics[scale=0.45]{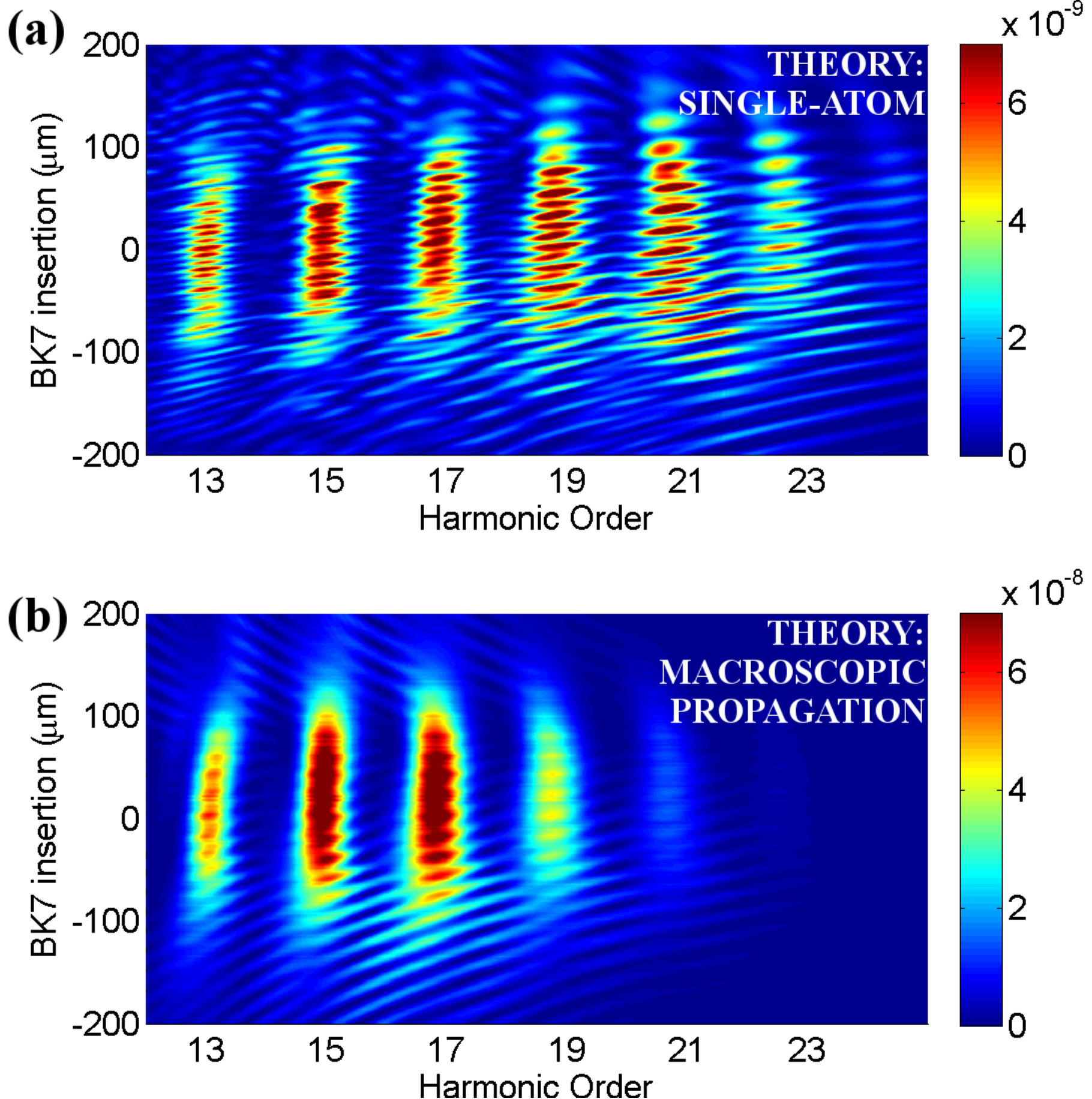}
	\caption{Simulated HHG from the single atom response (a) and including macroscopic propagation effects (b) for different chirps of the driving IR pulse. Chirp is scanned by the insertion of BK7, in a similar way as in the experiment. Zero insertion means a Fourier limit generating pulse ($4.3\ \mathrm{fs}$ in this case).}
	\label{fig2}
\end{figure}

Interestingly, the origin of the continuum behavior shown in Fig.$\ $\ref{fig1} can be found at both the microscopic and macroscopic levels. In Fig.$\ $\ref{fig2}(a) we represent the simulated microscopic single-atom spectra driven in argon versus BK7 insertion, where the driving laser pulse is modeled by a $\cos^2$ envelope function of $4.3\ \mathrm{fs}$ FWHM, with $2.5\times10^{14}\ \mathrm{W/cm}^2$ peak intensity and central wavelength $720\ \mathrm{nm}$. By changing the BK7 insertion, a standard harmonic spectrum structure is visible for positive and slightly negative chirp cases. However, for sufficiently negatively chirped input pulses, the HHG spectral structure changes, and spectral contributions appear out of the odd harmonic positions. Although the behavior of the single-atom CEP scan is similar to that observed experimentally, it does not fully reproduce our results.

Propagation effects alter further these spectral structures, as shown in Fig.$\ $\ref{fig2}(b), obtained for similar experimental focusing conditions. We use a Gaussian beam, whose waist at the entrance plane of the $40\ \mathrm{cm}$ focal length lens is $2.5\ \mathrm{mm}$, focused into a gas jet placed $2\ \mathrm{mm}$ before the focus position. The gas jet is modeled by a Gaussian distribution along the gas propagation direction with $1\ \mathrm{mm}$ FWHM and peak density of $10\ \mathrm{mbar}$. We show the harmonics as detected on-axis. Cutoff harmonics are lost during propagation \cite{lhuillier1993} and also odd harmonics show less chirp-dependent modulation than the single atom response in the positive chirp region. In the case of negative chirp, variations in the odd harmonic structure are still observed, but appear smoother than in the single atom case.  The modulated spectrum obtained in the single-atom response is filled when propagation is taken into account. 

\begin{figure}[h]
	\centering
	\includegraphics[scale=0.35]{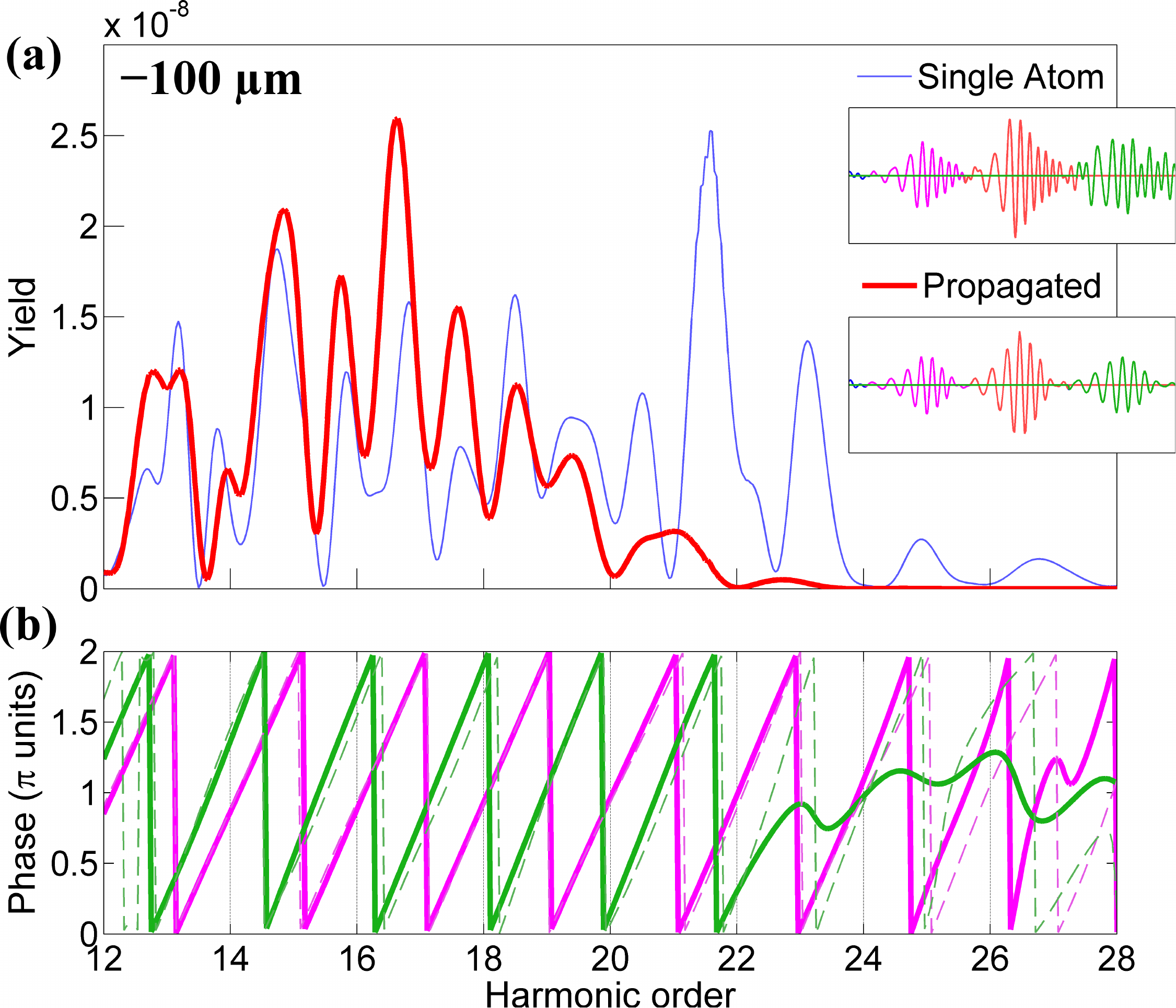}
	\caption{(a) Simulated XUV spectra when $100\ \mu\mathrm{m}$ of BK7 have been removed from the Fourier limit condition ($4.3\ \mathrm{fs}$ of pulse duration at Fourier limit). Blue thin line denotes single atom response while red thick line shows collective spectrum. Corresponding time evolution is shown as inset for both cases, single atom and propagated, denoting at different colors the main individual attosecond pulses. (b) Corresponding spectral phase differences between first and second attosecond pulses (thick violet line for propagated, thin dashed violet line for single atom), and second and third pulses (thick green line for propagated, thin dashed green line for single atom), are shown.}
	\label{fig3}
\end{figure}

\section{Discussion}

\begin{figure}[t]
	\centering
	\includegraphics[scale=0.35]{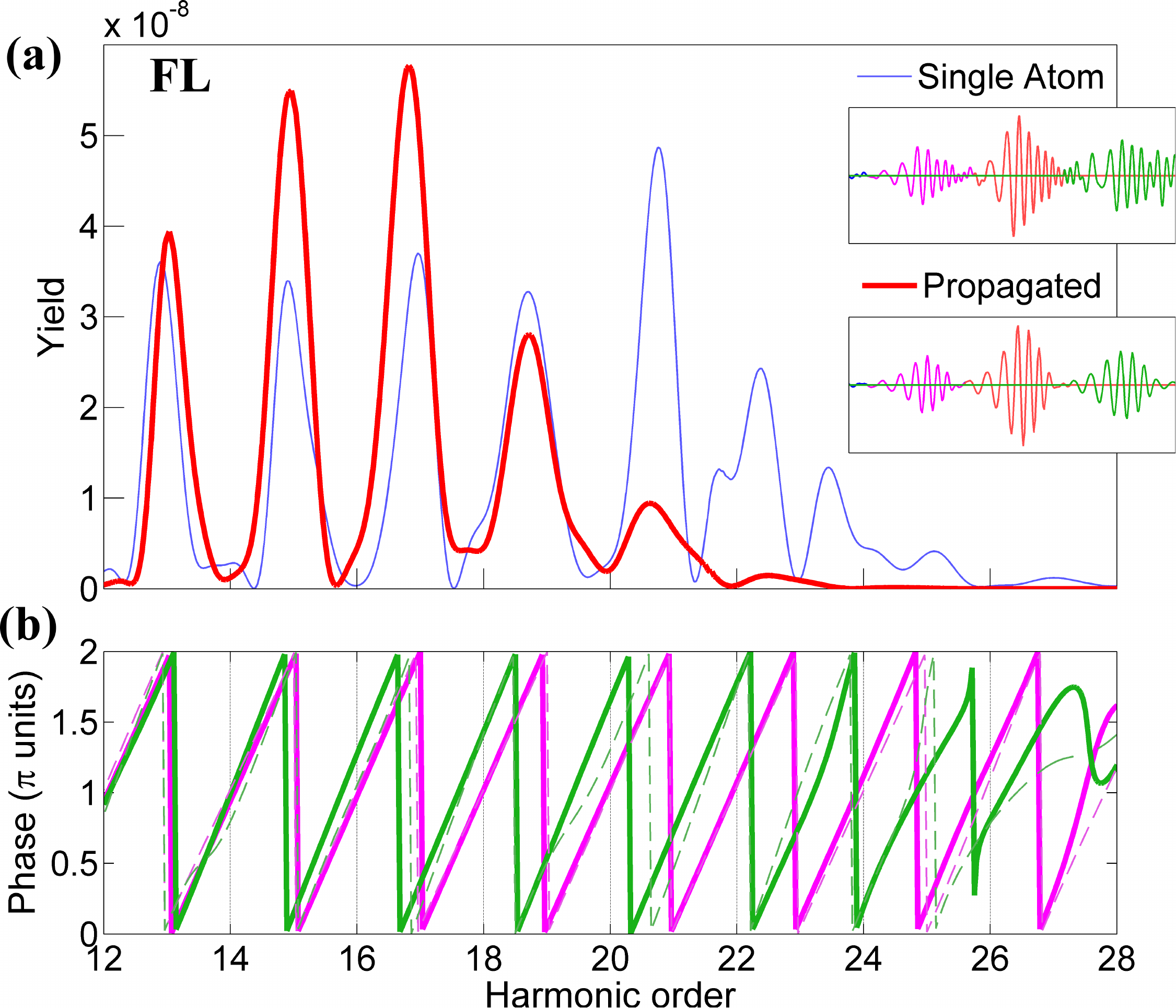}
	\caption{(a) Simulated XUV spectra for Fourier Limited (FL) driving pulses with a duration of $4\ \mathrm{fs}$. Blue thin line for single-atom and red thick line when the propagation is considered. In the insets, corresponding time profiles for single-atom and propagated simulations are shown . (b) Spectral phase differences between first and second attosecond pulses (thick violet line for propagated, dashed line for single atom), and second and third pulses (thick and dashed green lines).}
	\label{fig4}
\end{figure}

\begin{figure}[t]
	\centering
	\includegraphics[scale=0.35]{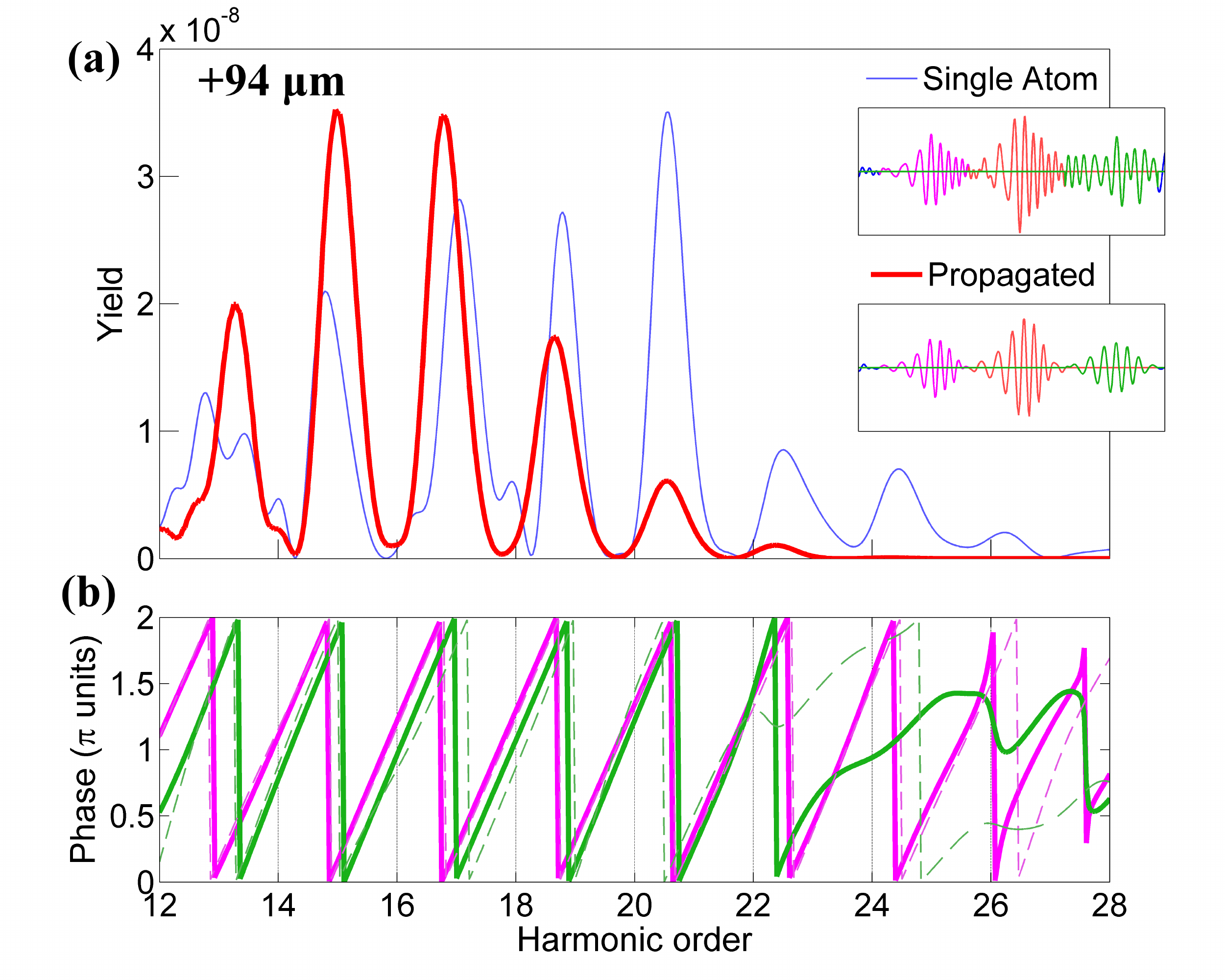}
	\caption{(a) Simulated XUV spectra for positively chirped driving pulses ($94\ \mu\mathrm{m}$ of BK7 added from the Fourier limit condition. Blue thin line for single-atom and red thick line for collective spectrum. In the insets, corresponding time profiles for single-atom and propagated simulations. (b) Spectral phase differences between first and second attosecond pulses (violet lines), and second and third pulses (green lines).}
	\label{fig5}
\end{figure}

For a further insight on the nature of the continua, we have analyzed the numerical simulations. For the sake of clarity, we compare a negatively chirped IR pulse, ($-100\ \mu\mathrm{m}$, Fig.$\ $\ref{fig3}), a Fourier limit pulse (Fig.$\ $\ref{fig4}) and a positively chirped pulse ($94\ \mu\mathrm{m}$, Fig.$\ $\ref{fig5}). In the harmonic emission spectrum [Figs.$\ $\ref{fig3}(a), \ref{fig4}(a) and \ref{fig5}(a)], blue lines denote single-atom response, while red lines represent calculations including propagation. In agreement with experiments, the negative chirp case yields a continuous spectrum [Fig.$\ $\ref{fig3}(a)] and spectra driven by unchirped or positively chirped IR fields show the usual HHG comb-like structure composed of odd harmonics [Figs.$\ $\ref{fig4}(a) and $\ $\ref{fig5}(a)]. The insets show the temporal structure for both the single-atom and the propagated cases. The different colors denote the main attosecond pulses within the train, amounting to three in this case.

The role of negative chirp in inducing changes over the harmonic structure at the single atom response level has been analyzed in HHG from solid targets \cite{borot2012}, observing a non-constant time separation between the attosecond pulses. This may lead to a Moir\'e pattern in the spectral domain, merging some harmonics and changing their wavelength but, in general, maintaining a peaked structure. In Ref.$\ $\cite{borot2012}, negative chirp in the IR driving field was identified as a major factor, since it enhances time delay emission differences, as demonstrated in Ref. \cite{varju2005} (in this case considering HHG from gas targets). In our case, this phenomenon is observed (see insets in Figs.$\ $\ref{fig3}, \ref{fig4} and \ref{fig5}), but notably the continuum is also associated to a particular phase difference among the individual attosecond pulses [Figs.$\ $\ref{fig3}(b), \ref{fig4}(b) and \ref{fig5}(b), where the violet (green) line represents the phase difference between the first and second (second and third) attosecond pulses]. The XUV spectra corresponding to the different chirp cases have been analyzed and described as the result of spectral interference of the single bursts, as suggested in Refs.$\ $\cite{ott2013,rudawski2015}.


For a negatively chirped driving field [Fig.$\ $\ref{fig3}(b)] the spectral phase difference between consecutive attosecond pulses within the pulse train [Fig.$\ $\ref{fig3}(a) inset] yields a continuum. Note that the continuum structure appears when the phase difference between the first (pink) and second (red) pulses differs from that between the second (red) and third (green) pulses. In other words, at wavelengths were dephasing of a pair of pulses is $\pi+2m\pi$ (\emph{m} being natural), a minimum is found in the spectrum due to destructive interference, and maxima arise from constructive interference when the dephasing is $2m\pi$. When the dephasings among the pulse pairs are similar [as it happens when using a Fourier limit or a positively chirped driving field, Figs.$\ $\ref{fig4}(b) and \ref{fig5}(b)], a sharp odd harmonic structure arises. Contrarily, when dephasings do not agree and even have a difference of $\pi$ rad [Fig.$\ $\ref{fig3}(b)], a more continuous structure emerges from interference [Fig.$\ $\ref{fig3}(a)]. Note that the dephasing is established at the single atom level [thin lines in Figs.$\ $\ref{fig3}(b), \ref{fig4}(b) and \ref{fig5}(b)], and is conserved when propagated (thick lines). The macrosopic addition and propagation of different XUV fields reinforces the obtention of the continuous structure.

\section{Conclusion}

To conclude, spectral continua have been observed in the medium energy range of high harmonic spectra, for adequate focusing and IR pulse conditions. Our simulations reveal that the origin of the continua can be understood not as the emission of an isolated attosecond pulse, but as the interference of several pulses in a train with different dephasing between consecutive pairs of pulses. The particular phase difference yielding the continuum structure can be obtained through adjustment of the focus position, IR pulse chirp and CEP, allowing a fine control of the resulting spectral interference pattern. Effects due to microscopic and macroscopic contributions are described and compared. From a practical point of view, this method enables obtaining continua in the XUV range without the need of pulse gating techniques, high intensities or single-cycle driving IR pulses.

\section*{Acknowledgements}

We acknowledge support by a Marie Curie International Outgoing Fellowship within the EU Seventh Framework Programme for Research and Technological Development, under REA Grant Agreement No. 328334. Support from Junta de Castilla y Le\'on (Project No. SA116U13), Spanish MINECO (Grant No. FIS2009-09522 and FIS2013-44174-P), and Centro de L\'aseres Pulsados, CLPU  is gratefully acknowledged. This work was partly supported by Grant PTDC/FIS/122511/2010, B. Alonso acknowledges support from Post-Doctoral fellowship SFRH/BPD/88424/2012 and H. Crespo acknowledges support from Sabbatical Leave Grant SFRH/BSAB/105974/2015, from Funda\c{c}\~{a}o para a Ci\^{e}ncia e Tecnologia, Portugal, co-funded by COMPETE and FEDER.


\end{document}